\documentclass[twocolumn,pra,aps]{revtex4}

\usepackage{graphicx}
\usepackage{color}
\usepackage[normalem]{ulem}
\usepackage{amsmath}

\def\duzomniejsze{<\kern-.7mm<}
\def\duzowieksze{>\kern-.7mm>}

\def\textbf#1{{\bf #1}}
\def\beq{\begin{equation}}
\def\eeq{\end{equation}}
\def\be{\begin{equation}}
\def\ee{\end{equation}}
\def\ben{\begin{eqnarray}}
\def\een{\end{eqnarray}}
\def\beqa{\begin{eqnarray}}
\def\eeqa{\end{eqnarray}}
\def\eea{\end{array}}
\def\bea{\begin{array}}
\newcommand{\bei}{\begin{itemize}}
\newcommand{\eei}{\end{itemize}}
\newcommand{\bee}{\begin{enumerate}}
\newcommand{\eee}{\end{enumerate}}

\def\tr{{\rm Tr}}

\def\>{\rangle}
\def\<{\langle}
\def\blacksquare{\vrule height 4pt width 3pt depth2pt}

\def\ot{\otimes}

\newtheorem{lemma}{Lemma}
\newtheorem{corrolary}{Corrolary}

\newtheorem{theorem}{Theorem}
\newtheorem{definition}{Definition}
\newtheorem{observation}{Observation}

\def\bed{\begin{definition}}
\def\eed{\end{definition}}
\def\bel{\begin{lemma}}
\def\eel{\end{lemma}}
\def\bet{\begin{theorem}}
\def\eet{\end{theorem}}

\begin{document}

\title{No-broadcasting of non-signalling boxes via operations which transform local boxes into local ones}

\begin{abstract}
We deal with families of probability distributions satisfying non-signalling condition, called non-signalling boxes and
consider a class of operations that transform local boxes into local ones 
(the one that admit LHV model). We prove that any operation from this class
cannot broadcast a bipartite non-local box with 2 binary inputs and outputs. We consider a function called
anti-Robustness which can not decrease under these operations. The proof reduces to showing
that anti-Robustness would decrease after broadcasting.
\end{abstract}

\author{P. Joshi$^{1,2}$, A. Grudka$^3$, K. Horodecki$^{2,4}$,  M. Horodecki$^{1,2}$, P. Horodecki$^{2,5}$ and R. Horodecki$^{1,2}$}
\affiliation{$^1$Faculty of Mathematics,Physics and Informatics, University of Gda\'nsk, 80--952 Gda\'nsk,Poland}
\affiliation{$^2$National Quantum Information Center of Gda\'nsk, 81--824 Sopot, Poland}
\affiliation{$^3$Faculty of Physics, Adam Mickiewicz University, 61-614 Pozna\'n, Poland }
\affiliation{$^4$Institute of Informatics, University of Gda\'nsk, 80--952 Gda\'nsk, Poland}
\affiliation{$^5$Technical University of Gda\'nsk, 80--233 Gda\'nsk, Poland}


\maketitle
\section{introduction}
Given a quantum bipartite state and a set of measurements on its both subsystems, one ends up with a family of probability distributions obtained from these measurement on the quantum state. Such a family can have interesting features, e.g. can violate some of the Bell inequalities \cite{Bell}. Moreover such a family satisfies the so-called non-signalling condition: change of measurement by one party can not change statistics of the other party. One can then ask after Popescu and Rohrlich \cite{PR}, if any set of non-signalling distributions (called a box),  can be reproduced by measurement on quantum state. The answer is no, and the proof is given by the fact, that certain (called Popescu-Rohrlich) boxes violate CHSH inequality up to 4, while maximal 
violation via measurements on quantum states of this inequality is due to Cirel'son's limit $2\sqrt{2}$ \cite{Cirel}. 

Since this discovery by Popescu and Rohrlich, non-signalling 
boxes have been treated as a resource in different contexts \cite{Barret-Roberts}. In particular it has been shown, that they bear analogous features to those of entangled states \cite{Masanes-NStheor} such as non-shareability \cite{Masanes-NStheor}, monogamy of correlations \cite{Pawlowski-Brukner}, offering secret key \cite{Ekert91,BHK_Bell_key,Bell-security} which lead to the so called device independent security (see \cite{Hanggi-phd} and references therein). The distillation of PR-boxes and cost of non-locality has attracted recently much attention as well \cite{BrunnerSkrzypczyk,Allcocketal2009,Forster,Brunneretal2011}, as an analogue of distillation of entanglement.

Another context in which non-signalling principle was considered, are the well known no-goes of quantum theory: no-cloning \cite{WoottersZ-cloning} and no-broadcasting \cite{no-broadcast}. The first states that there is no universal machine which given an unknown input produces its copies, while the second is stronger: it states that there is no universal machine that given an unknown state $\rho$ produces a state whose subsystems are in state $\rho$. Analogous results for non-signalling boxes were shown in \cite{GPT-noclon-nobroad}.

There is also a bipartite version of no-cloning and no-broadcasting theorems. In case of bipartite quantum states one requires that the input state of machine is {\it known}, but the operations which machine uses are not all quantum operations but {\it local operations} \cite{Piani-no-broad,Hor-Sen-no-clon} (see also \cite{Piani2009-broadcast}). It was shown later, that this kind of no-broadcasting is equivalent to the previous mentioned one, (with general operations) in \cite{SLuo,SLuoWSun}. 

In this article, we consider a variant of 'local' broadcasting of bipartite non-signalling boxes with 2 binary inputs and outputs. (We will represent all such boxes as $2\times 2$ for our convenience). Namely we assume that the input box is known, and it is processed by locality preserving operations. By locality preserving operations we mean here the ones that transform local boxes (those with local hidden variable model) into local ones. We show that any $2\times 2$ non-local box cannot be broadcast in two copies (which excludes broadcasting in arbitrary n-copies), and prove it using idea of monotones, in analogy to entanglement theory.
Namely we introduce monotone called anti-Robustness, which cannot decrease under locality preserving operations. For a related entanglement monotone see \cite{VidalT1998-robustness}. 
We then show that if broadcast were possible, anti-Robustness would increase under locality preserving operations, which gives desired contradiction. Our proof has two main parts: we first show this for states which are mixture of PR and anti-PR boxes and then show that broadcasting of any other non-local boxes implies broadcasting of the latter case. By symmetry, we then extend the argument to all $2\times 2$ boxes.  We begin however with analogous question in quantum case: can one broadcast a quantum bipartite state which is entangled by means of operations that transform separable states into separable ones, and answer in negative to this question in section \ref{sec:quantum}. The main tools are introduced in section \ref{sec:line} and \ref{subsec:twirl}. The two parts of the proof of main result are in sections \ref{sec:lineproof} and \ref{subsec:all}.
Appendix contains proofs of some needed facts.

\section{No-broadcasting in quantum case}
\label{sec:quantum}
In this section we show that any entangled states can not be broadcast, which is in 
fact an immediate implication of known facts from entanglement theory.

Let $\rho_{AB}$ be a state and $\Lambda$ be a broadcast map which maps $\rho_{AB}$ to $\rho_{ABA'B'}$ i.e.
\be \label{eq1} \Lambda(\rho_{AB}) = \rho_{ABA'B'}, \ee
and 
\be Tr_{AB}\left ( \rho_{ABA'B'}\right ) = Tr_{A'B'} \left ( \rho_{ABA'B'} \right )= \rho_{AB}. \ee

We show now that such a map does not exist if $\rho_{AB}$ is entangled and it preserves the set of separable states.

To this end, consider first cloning of known bipartite state by LOCC operations which is a smaller class of operations
than the separability preserving ones. Cloning is nothing but creating independent copies i.e. in (\ref{eq1}) demanding $ \Lambda(\rho_{AB}) = \rho_{AB}^{\otimes2}$.
Such a problem was considered in \cite{dual} 
and entanglement measure defined by quality of cloning was suggested. 
Indeed, by LOCC one can clone any separable state, so if the quality of 
cloning is not perfect the state must be entangled.
At that time it was not known whether any entangled state can be cloned. 
Note that, if we knew an entanglement measure $E$ (a function that does not increase under LOCC), that for any entangled state 
\be
E(\rho_{AB}\ot\rho_{A'B'})>E(\rho_{AB}),
\ee
this would imply impossibility of cloning entangled state by LOCC.  
Indeed, by cloning we would increase the measure $E$, which is impossible by LOCC. 
Consider now broadcasting by LOCC. Again, if we knew entanglement measure $E$ 
which satisfies even more, namely 
\be
E(\rho_{ABA'B'})>E(\rho_{AB}),
\label{eq:super}
\ee
for any entangled state $\rho_{AB}$ 
and any state $\rho_{ABA'B'}$ being a broadcast copy of $\rho_{AB}$,
i.e. $\tr_{AB} (\rho_{ABA'B'})=\rho_{AB}$ and $\tr_{A'B'} (\rho_{ABA'B'})=\rho_{AB}$,
then broadcasting of known entangled state  by LOCC would be impossible.
Indeed, like in cloning case, by broadcasting, we would increase the measure $E$, 
which is impossible by LOCC.

Such a measure is actually known. Namely, in \cite{YangHHS2005-cost} it was shown that 
entanglement of formation satisfies this equation, and as a corollary, it was obtained that 
cloning of arbitrary entangled state by LOCC is impossible (and also broadcasting, 
as we now see). 

One can strengthen the result by referring to a later analogous 
result by Marco Piani \cite{Piani2009-relent},
who showed that relative entropy of entanglement satisfies 
equation (\ref{eq:super}) too for any entangled state. 
Now, since relative entropy of entanglement does not increase 
after action of arbitrary operations which preserve the set of separable states \cite{BrandaoPlenio2007-separable-maps} (called 
non-entangling operations), 
we obtain 
{\corrolary
Arbitrary entangled state cannot be broadcast by non-entangling operations. 
}

Finally, let us mention, that broadcasting by means of local operations 
was also considered, and it was shown in \cite{Piani-no-broad}
that a state can be broadcast by means of local operations only 
when the state is classical (i.e. it is a state of two classical registers).

\section {Statement of the broadcasting problem in box scenario}
\label{sec:statement}
By box $X$ we mean a family of probability distributions that have support on Cartesian product of spaces $\Omega_{A}\times \Omega_{B}$.
Each of the spaces may contain (the same number of) $n$ systems i.e. it may be a product of spaces $\Omega_{A_1}...\Omega_{A_n}$. We will consider only boxes that satisfy certain non-signalling conditions.
To specify this we need to define a general non-signalling \cite{Barrett-GPT,Barret-Roberts} condition between some partitions of systems.

{\definition \label{def:NS}Consider a box of some number of systems $n+m$ and its partition into two sets: $A_1,...,A_n$ and $B_1,...,B_m$. A box on these systems
given by probability table $P(\bf{a},\bf{b}|\bf{x},\bf{y})$ is non-signalling in cut $A_1,...,A_n$ and $B_1,...,B_{m}$
if the following two conditions are satisfied:
\ben
\forall_{\bf{a},\bf{x},\bf{x},\bf{y}'} \sum_{\bf{b}}P({\bf{a}},{\bf{b}}|{\bf{x}},{\bf{y}}) = \sum_{\bf{b}}P(\bf{a},\bf{b}|\bf{x},\bf{y}')\\
\forall_{{\bf{b}},\bf{x},{\bf{x}'},{\bf{y}}} \sum_{\bf{a}}P({\bf{a}},{\bf{b}}|{\bf{x}},{\bf{y}}) = \sum_{\bf{a}}P(\bf{a},{\bf{b}}|\bf{x}',\bf{y})
\een

If the first condtion is satisfied, we denote it as 
\be A_1,...,A_n\not\hspace{-1.3mm}{\rightarrow} B_1,...,B_{m},\nonumber
\ee if the second we write
\be
B_1,...,B_{m}\not\hspace{-1.3mm}{\rightarrow} A_1,...,A_n,\nonumber
\ee and if both:
\be
A_1,...,A_n\not\hspace{-1.3mm}{\leftrightarrow} B_1,...,B_{m}\nonumber .
\ee
We say that a box of systems $A_1,...,A_n,B_1,...,B_{m}$ is fully non-signaling if for any subset of systems $A^IB^J \equiv A_{i_1},...,A_{i_k}B_{j_1},...,B_{j_l}$ with $I\equiv\{i_1,...,i_k\}\subseteq N \equiv\{1,...,n\}$ and $J\equiv\{j_1,....,j_l\}\subseteq M \equiv\{1,...,m\}$ such that not both I and J are empty, there is 
\be A^IB^J\not\hspace{-1.3mm}{\leftrightarrow}A^{N-I}B^{M-J}.
\ee
}

By locally realistic box we mean the following ones:
{\definition Locally realistic box of $2n$ systems $A_1,...,A_n,B_1,...,B_n$ is defined as 
\be
\sum_\lambda p(\lambda)P{(\bf{a}|\bf{x})}_{A_1,...,A_n}^{(\lambda)}\otimes P{(\bf {b}|\bf {y})}_{B_1,...,B_n}^{(\lambda)}
\label{eq:local}
\ee
for some probability distribution $p(\lambda)$, where we assume that 
boxes $P{(\bf{a}|\bf{x})}_{A_1,...,A_n}^{(\lambda)}$ and 
$P{(\bf{b}|\bf{y})}_{B_1,...,B_n}^{(\lambda)}$ are fully non-signaling. 
The set of all such boxes we denote as $LR_{ns}$. All boxes that are fully non-signaling
but do not satisfy the condition (\ref{eq:local}), are called non-$LR_{ns}$.
}

Having defined relevant classes of boxes, we can define relevant class of operations.
We consider a family $\cal L$ of operations $\Lambda$ on a box shared between Alice and Bob, which preserve locality, as defined below.

{\definition An operation $\Lambda$ is called locality preserving if it satisfies the following conditions:}

(i) {\it validity} i.e. transforms boxes into boxes.

(ii) {\it linearity} i.e. for each mixture $X = p P + (1-p) Q$, there is $\Lambda(X) = p \Lambda(P) + (1-p) \Lambda(Q)$

(iii) {\it locality preserving } i.e. transforms boxes from $LR_{ns}$ into boxes from $LR_{ns}$.

(iv) transforms a {\it fully non-signalling} box as defined in the Def(\ref{def:NS}) into fully non-signalling one.
\newline The set of all such operations we denote as $\cal L$.

The problem of broadcasting is then {\it if there exists $\Lambda \in {\cal L}$ which makes a 2-copy broadcast of a $2\times 2$ box P} i.e.
\be
\Lambda(P_{AB}) = P^{(2)}_{ABA'B'},
\ee
where $ P^{(2)}$ is arbitrary box with 4 inputs and 4 outputs whose marginal boxes on AB and A'B' are again box P. More generally,
one can ask if there exists $\Lambda$ which produces $n$ broadcast copies,  which is a problem of n-copy broadcast. The impossibility of 2-copy broadcast implies the impossibility of n-copy broadcast. Hence it is enough to deal with 2-copy broadcast. From now on by broadcast we mean 2-copy broadcast.

\section{No-broadcasting theorem in box scenario}
\label{sec:line}

Our approach is like in entanglement theory (see in this context \cite{Short}). We pick up a monotone and show that it could be smaller after broadcasting which is not possible. Our monotone
will be {\it anti-Robustness of a box}, defined as follows:

{\definition
Let A be a NS and non-$LR_{ns}$ box. 
Anti-Robustness of A, $\bar{R}(A)$ is defined as,
\be
\bar{R}(A)=\max_{X}\{q| qA+(1-q)X \in LR_{ns} \},
\ee
where $X$ is arbitrary NS box.
}

The name anti-Robustness comes from the fact that if a given $q$ is anti-Robustness of some box A, then
$1-q$ is minimal weight with which one needs to admix some box X to make A local, i.e. $1-q$ reports how
'robust' is A against admixing of some other boxes in terms of non-locality. It is analogous to the robustness quantity defined in\cite{VidalT1998-robustness}. It would be measure of non-locality in the same way as the Robustness of entanglement is measure of entanglement (cf.  other measures of nonlocality \cite{Brunneretal2011,Fitzi}.

{\observation ${\bar{R}}$ is non decreasing under locality preserving operations.
\label{obs:monot}
}

{\it Proof}.- Let us fix an arbitrary NS box A. Let $\Lambda$ be linear operation taking $LR_{ns}$ boxes into $LR_{ns}$ boxes. Let also $q_0$ be the value of ${\bar{R}}(A)$.
Then there exists box $X$ such that $q_0A + (1-q_0)X \in LR_{ns}$. Let us apply $\Lambda$ to $q_0A + (1-q_0)X$ by linearity of $\Lambda$ it reads
$q_0\Lambda(A)+(1-q_0)\Lambda(X)$, and by its locality, this box is LR, hence $q_0$ is a candidate for value ${\bar{R}}(\Lambda(A))$, but by definition
the latter can be at most higher, hence proving ${\bar{R}}(A)\leq {\bar{R}}(\Lambda(A))$.\blacksquare

In the following, we will need also a technical property of anti-Robustness, that can be viewed as connectivity: if it is attained at $q$, it could be 
attained at all $p<q$:

{\observation If there exists $L=qA + (1-q)X$ with $q > p$, then there exists also a box $X'$ such that $L= pA + (1-p)X'$.
\label{obs:continuity}
} 

{\it Proof}.- The proof is straightforward with $X'= (q-p)/(1-p)A + (1-q)/(1-p)X$.\blacksquare

\subsection{Extremal non-local boxes, twirlings and CHSH quantities}
\label{subsec:twirl}
We will use numerously the operation of twirling of a box \cite{Nonsig_theories,Short}, which maps all boxes into a smaller subset of boxes. In what follows we will consider 4 such twirlings and show that they preserve corresponding CHSH quantities.  

Let us recall the geometry of two parties 2 binary inputs and outputs NS boxes. The set of such boxes forms an 8 dimensional polytope with 24 vertices~\cite{Barret-Roberts}. 16 of these are deterministic boxes which span the set of $LR_{ns}$ boxes that satisfy all the CHSH inequalities \cite{Barret-Roberts} $-2\leq \beta_{rst}(P) \leq 2$ for all choices of $r, s$ and $t$ with:
\ben
\beta_{rst}(P) & \equiv & (-1)^t\<00\>+(-1)^{s+t}\<01\>+(-1)^{r+t}\<10\> \nonumber\\
&& + (-1)^{r+s+t+1}\<11\>,
\label{eq:betarst}
\een
where $\<ij\>=P(a=b|ij)-P(a\neq b|ij)$ and r, s, t takes values either 0 or 1. The rest of 8 vertices are equivalent to PR-boxes, which are defined as follows :

\be B_{rst}(ab|xy) = \left\{ \begin{array}{ll}
         1/2 & \mbox{if $a\oplus b = xy\oplus rx \oplus sy \oplus t$} \\
        0 & \mbox{else}.\end{array} \right.
\ee

We describe now 4 twirling operations $\tau_{rs}$, and show that they preserve corresponding CHSH quantities. The twirling $\tau_{00}$ is introduced in \cite{Nonsig_theories,Short}.

{\definition A twirling operation $\tau_{rs}$ is defined by flipping randomly 3 bits $\delta,\gamma,\theta$ and applying the following transformation to a $2\times 2$ box $P(a,b|x,y)$:
\ben
x &\rightarrow & x \oplus \delta \nonumber \\
y &\rightarrow &y \oplus \gamma \nonumber \\
a &\rightarrow &a \oplus \gamma x \oplus \delta\gamma \oplus \theta \oplus s\gamma \nonumber \\
b &\rightarrow &b \oplus \delta y \oplus \theta \oplus r\delta \nonumber \\
\een
\label{def:twirling}
}

We then make the following observation, which is easy to check:

{\observation Twirling $\tau_{rs}$ maps all $2\times 2$ boxes into line $pB_{rst} + (1-p)B_{rs\bar{t}}$; 
$\tau_{rs}(B_{rst})=B_{rst}$ and $\tau_{rs}(B_{rs\bar{t}})=B_{rs\bar{t}}$ where $\bar{t}$ denotes binary negation of $t$.
\label{obs:twirling}
}

We are ready to show that twirling preserves appropriate CHSH quantity, which is formulated in lemma below:

{\lemma The CHSH quantities $\beta_{rst}$ 
satisfy:
\be
\beta_{rst}(P) = \beta_{rst}(\tau_{rs}(P)).
\ee 
\label{lem:CHSH}
}

{\it Proof}.- It is straightforward to check that 
$\beta_{rst}(P) = \<2(B_{rst} - B_{rs{\bar{t}}})|P\>$, where $\<.|.\>$ denotes Euclidean scalar product and 
hence there is 
 \begin{align}
\beta_{rst}(\tau_{rs}(P))= \<2(B_{rst} - B_{rs{\bar{t}}})|\tau_{rs}(P)\>&=\nonumber\\ = \sum_i q_i \<2(B_{rst} - B_{rs{\bar{t}}})|\pi_i P\>.
\end{align}
Here we use the fact that each twirling is a mixture of some permutations $\pi_i$. We have 
then 
 \begin{align}
\beta_{rst}(\tau_{rs}(P)) = \sum_i q_i \<\pi_i 2(B_{rst} - B_{rs{\bar{t}}})| P\>&=\nonumber\\=\<\tau_{rs}(2(B_{rst} - B_{rs{\bar{t}}}))|P\>
\end{align}

which ends the proof, since $\tau_{rs}(B_{rst})= B_{rst}$ and $\tau_{rs}(B_{rs\bar{t}})= B_{rs\bar{t}}$ by
observation \ref{obs:twirling}.\blacksquare



\subsection{No-broadcasting for mixtures of PR and anti-PR box}
\label{sec:lineproof}
Here we show, for a subclass $\{B_\alpha\}$ of non-$LR_{ns}$ boxes that they cannot be broadcast. These boxes are a family of convex combinations of $B_{000}\equiv B$ (PR-box) and $B_{001}\equiv \tilde{B}$ (anti-PR) boxes i.e.
\be \label{eq:Balpha}
B_{\alpha} = \alpha B + \left (1 - \alpha \right ) \tilde{B}, 
\ee
where $\alpha \in [1,{3\over 4})$,
such that when $\alpha = 1 \Rightarrow  B_{\alpha} = B$ (PR box). For $\alpha=\frac{3}{4}$, $B_\alpha$ becomes an LR box (say $K$)
\be
K=\frac{3}{4}B+\frac{1}{4}\tilde{B}
\label{eq:localbox}\ee

One can express $K$ in terms of $B_{\alpha}$  and $\tilde{B}$ as follows:
\be
K= p_{\alpha} B_{\alpha} + (1- p_{\alpha}) \tilde{B}, 
\label{eq:convx1} \ee 
where $p_{\alpha} = {3\over 4\alpha}$.
We show that for $B_{\alpha}$ in (\ref{eq:Balpha}) broadcasting is not possible. Only when $B_{\alpha}=K$, it turns out that broadcasting can be possible which is known fact for $LR_{ns}$ boxes.

Before passing to the main results there is an important observation required in the main proof:

{\observation ${\bar{R}}(B_{\alpha}) = p_{\alpha} $. \label{obs:palpha}}

{\it Proof}:- Consider, two boxes $C$ and $X$ such that $\beta(X) \in [-4,\beta(C)]$ and $\beta(C)\in [-2,2]$. These two boxes lie on the line of convex combination of $B$ and $B'$. Let us write $C$ as follows,
\be
C=qB_\alpha +(1-q)X
\label{eq:antiR} \ee
Taking $\beta$ values of (\ref{eq:antiR}) and  rearranging terms we get
\be
q=\frac{\beta(C)-\beta(X)}{\beta(B_\alpha)-\beta(X)}
\ee
By the definition of anti-Robustness, $\bar{R}(B_\alpha)$ is equal to maximum of $q$ over all $X$ such that (\ref{eq:antiR}) holds. So to maximize $q$ we need to have maximum value of $\beta(C)=2$.
\be
q_{max} = \frac{2-\beta(X)}{8\alpha-4 -\beta(X)}
\ee
We find that for $\alpha \in ({3\over4},1]$, $q_{max}={3\over{4\alpha}}$.
Hence, ${\bar{R}}(B_{\alpha}) = {3\over{4\alpha}}=p_{\alpha}$.

{\theorem For any broadcast copy $\hat{B}_{\alpha}$ of $B_{\alpha}$, where $\alpha \in (\frac{3}{4},1]$, there is 
\be 
{\bar{R}}(B_{\alpha}) > {\bar{R}}(\hat{B}_{\alpha}).
\label{eq:non-inc-online}
\ee
}

{\it Proof}.- Suppose we can broadcast $B_{\alpha}$ and consider $\hat{B}_{\alpha}$ be the broadcast of it. 
Let $\Lambda$ be the operation that achieves broadcast. Since it transforms $LR_{ns}$ boxes into $LR_{ns}$ boxes, if applied to $K$ it will transform it into some box L $\in LR_{ns}$. 
Thus using (\ref{eq:convx1}) we would obtain:
\be
L=p_{\alpha}\hat{B}_{\alpha} + (1- p_{\alpha}) X.
\label{eq:maineq}
\ee 
We will show however in lemma \ref{lem:non-exist} below, that for any broadcast copy $\hat{B}_{\alpha}$, any $LR_{ns}$ box L and any box X the above equality does not hold. Now, if there does not exist an $LR_{ns}$ box L satisfying the above equality, by definition of anti-Robustness $p_{\alpha} (= {\bar{R}}(B_{\alpha}))$ can not be anti-Robustness of 
$\hat{B}_{\alpha}$. The latter can not be also higher than $p_{\alpha}$ or otherwise (\ref{eq:maineq}) would be satisfied, because we have the observation \ref{obs:continuity} which ends the proof.\blacksquare

{\corrolary The boxes $B_{\alpha}$ for $\alpha \in ({3\over 4},1]$ are not broadcastable.
}

{\it Proof}.- The proof follows from the above theorem and monotonicity of anti-Robustness (observation \ref{obs:monot}).\blacksquare

We can proceed now with the proof of crucial lemma mentioned in the proof of the theorem above:

{\lemma For any broadcast copy $\hat{B}_{\alpha}$ of $B_\alpha$with $\alpha \in ({3\over 4},1]$, any $LR_{ns}$ box L and any NS box X the equality  
\be
L=p_{\alpha}\hat{B}_{\alpha} + (1- p_{\alpha}) X
\label{eq:not-exist}
\ee 
does not hold.
\label{lem:non-exist}
} 

{\it Proof}.-
We define two random variables $C_1$ and $C_2$ in the following way: pick randomly independent 2 bits and according to it's values choose inputs x and y for the first box. Then compare the outputs a and b. If outputs satisfy the condition $a\oplus b=xy$, the value of $C_1$ is set to 4 otherwise it is set to -4. It is straightforward to check that the average value of this random variable on a box equals the CHSH quantity $\beta_{000}$ of the first box. We define $C_2$ in the same way on second box.

We will be interested now in joint probability distributions of the variables $(C_1,C_2)$ applied to boxes L, $\hat{B}_{\alpha}$ and $X$. 
Note that this transformation of mapping the box to a probability distribution is linear.
With a box we associate a corresponding probability distribution of $(C_1,C_2)$:

\be
L \rightarrow \{p'_{11} , p'_{12}, p'_{21}, p'_{22}\},
\ee
with $p'_{ij}=P(C_1=(-1)^{i+1}4,C_2=(-1)^{j+1}4)$, similarly for a broadcast copy:
\be
\hat{B}_{\alpha} \rightarrow \{\tilde{p}_{11}, \tilde{p}_{12},\tilde{p}_{21},\tilde{p}_{22}\},
\ee
and the $X$:
\be
X \rightarrow \{p''_{11},p''_{12},p''_{21},p''_{22}\}.
\ee
If there exists L and $X$ such that (\ref{eq:not-exist}) holds, then there also exists the one which is permutationally invariant w.r.t. to copies i.e. symmetric, because $\hat{B}_{\alpha}$ is such. Hence without loss of generality, we can assume that distribution $L$ is symmetric, therefore we have  $p'_{12}$=$p'_{21}$ and $\tilde{p}_{12}$ = $\tilde{p}_{21}$.

Now, by assumption, the box $L$ is $LR_{ns}$. Hence, if we perform operation defined by the random variable $C_1$ on first copy, given we observe value 4, the second copy is also an $LR_{ns}$ box: it is a mixture of induced product boxes on the second system given input $i,j$ and output $a,b$ on the first system \cite{KH-discrim}. Analogous property holds if we ask how the second system looks like given the $C_1$ had value $-4$. 
Thus the CHSH inequality of conditional box on second system must hold in both cases. Recall that the average value of $C_2$ on a box equals the CHSH quantity $\beta_{000}$ of this box, thus if the measured copy results 4 for $C_1$, the CHSH inequality $-2\leq \beta_{000}\leq 2$ for the second copy is
\be
-2 \leq \frac{4p'_{11} + (-4)p'_{12}}{p'_{11} + p'_{12}} \leq 2,
\ee
and if the outcome is $-4$ then CHSH inequality is 
\be
-2 \leq \frac{4p'_{21} + (-4)p'_{22}}{p'_{21} + p'_{22}} \leq 2.
\ee
Using the fact that $p'_{12}$=$p'_{21}$, $\tilde{p}_{12}$ = $\tilde{p}_{21}$ and normalization condition, we rewrite the above inequalities in simplified form as follows
\be
0 \leq 6 p'_{11} - 2 p'_{12},
\label{eq:first}
\ee
\be
2 p'_{11} - 6 p'_{12} \leq 0,
\ee
\be
0 \leq 2 p'_{11} + 10 p'_{12} - 2,
\label{eq:line1}
\ee
\be
6 p'_{11} + 14 p'_{12} - 6 \leq 0.
\label{eq:line2}
\ee
These inequalities give an $LR_{ns}$ polytope.\newline
\begin{figure}[h!]
  \centering
      \includegraphics[width=6cm]{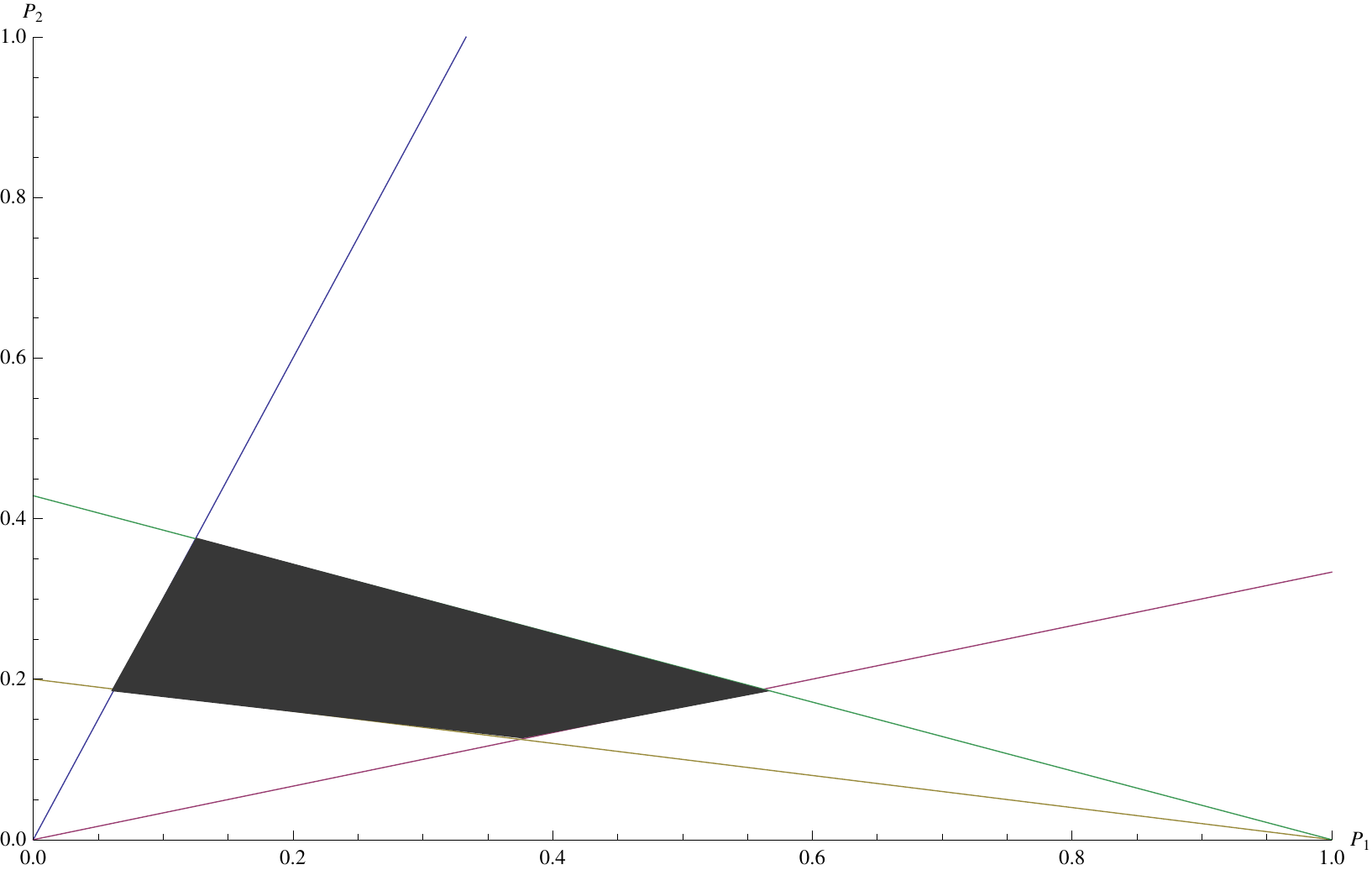}
  \caption{A set of pairs of parameters $(p_{11}',p_{12}')$ of box L (shaded region), satisfying locality conditions (\ref{eq:first})-(\ref{eq:line2}).}
\end{figure}
\newline
Let us consider constraints on $\hat{B}_{\alpha}$, i) normalisation condition ii) symmetry of broadcast i.e. $\tilde{p}_{12} = \tilde{p}_{21}$ iii) if we trace out one copy of the broadcast copy then the second copy has to be $B_{\alpha}$ and hence 
\be
\tilde{p}_{11} + \tilde{p}_{21} = \alpha,
\label{eq:line}
\ee
as $\alpha$ is probability of obtaining $4$ (i.e. value 4 of $C_1$) on $B_{\alpha}$ box.
We can rewrite eq.(\ref{eq:not-exist}) as,
\be 
\label{eq:eq1}
L - p_{\alpha}\hat{B}_{\alpha} = (1- p_{\alpha}) X.
\ee
Thanks to linearity of the map giving $(C_1,C_2)$ for a box, the same relation holds for the related probability distributions $\{p'_{ij}\}$, $\{\tilde{p}_{ij}\}$ and $\{p''_{ij}\}$.
Since $X$ is a box, the distribution $\{p''_{ij}\}$ should always have positive coefficients. We check if for any $B_{\alpha}$ it becomes negative.
So, now we have a complete set of $LR_{ns}$ boxes mapped on the shaded region as shown in fig.1 let us denote it $S_1$. The image under mapping to distribution $\{\tilde{p}_{ij}\}$ of one copy of $\hat{B}_{\alpha}$ is nothing but a straight line given by the eq. (\ref{eq:line}). We draw this straight line scaled by the factor $p_{\alpha}$, and denote resulting set of points as $S_2$. Interestingly, for any $\alpha \in [\frac{3}{4},1]$, we get the same line, because $(p_{\alpha} \tilde{p}_{11}, p_{\alpha}\alpha-p_{\alpha} \tilde{p}_{11},)=(p_{\alpha} \tilde{p}_{11}, \frac{3}{4}-p_{\alpha} \tilde{p}_{11})$. Changing value of  $\alpha$, simply shifts points on the line since slope of the line $p_{\alpha}\hat{B}_{\alpha}$ is the same for all $\alpha$.
\begin{figure}[h!]
  \centering
      \includegraphics[width=6cm]{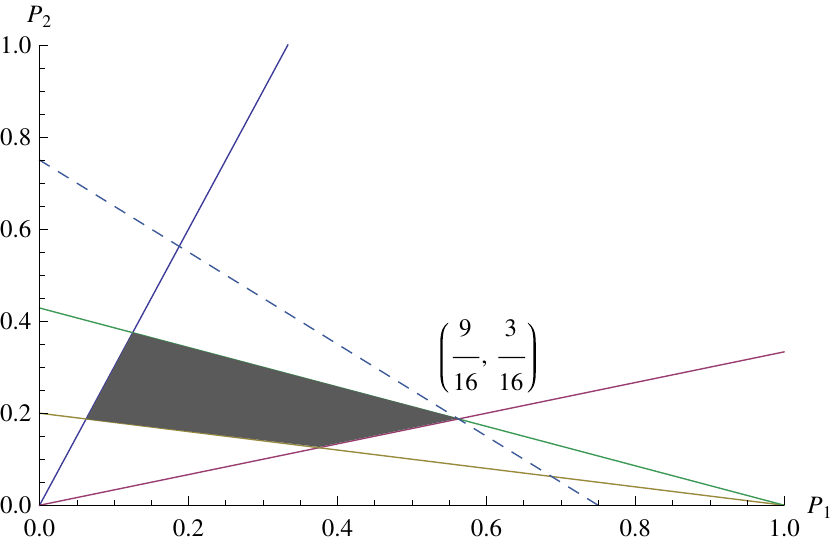}
  \caption{A set of pairs of parameters $(p_{11}',p_{12}')$ of box L, satisfying locality conditions (\ref{eq:first})-(\ref{eq:line2}) (shaded region) and the set
  of the parameters $(\tilde{p}_{11},\tilde{p}_{21})$ of box $\hat{B}_{\alpha}$ scaled by $p_{\alpha}$ (dashed line).}
\end{figure}
\newline In fig.2, we draw points of $S_1$ and $S_2$ and find that they only intersect at a point ($\frac{9}{16},\frac{3}{16}$). For all other points, $(1-p_{\alpha})X$ turns out to be negative. At the intersection, since $X$ is normalised and LHS in (\ref{eq:eq1}) is zero, $p_{\alpha}$ should be 1. Therefore, $\alpha=\frac{3}{4}$. This intersection point precisely corresponds to the case when $B_{\alpha} = K$, which we know can be broadcast. Hence for any $\alpha \in (\frac{3}{4}, 1]$ the eq. (\ref{eq:eq1}) does not hold with both positive $X$. This ends proof of the lemma \ref{lem:non-exist}. \blacksquare

\subsection{ General Case - No-broadcasting for all $2\times 2$ non-$LR_{ns}$ boxes }
\label{subsec:all}
In this section we show no-broadcasting for all $2\times 2$ non-$LR_{ns}$ boxes. 
To this end we will need the following crucial lemma, proved in Appendix.

{\lemma For any r,s,t in $\{0 , 1\}$, and any box $P_{AB}$ satisfying $\beta_{rst}(P_{AB})\geq 2$ there is 
\be {\bar{R}}(P_{AB}) = {\bar{R}}(\tau_{rs}(P_{AB})), \ee 
where $\tau_{rs}$ is twirling given in def \ref{def:twirling}.
\label{lem:tau-commutes}
}

We are ready to state our main result:

{\theorem Any non-$LR_{ns}$ box in $2\times 2$ can not be broadcast.
}

{\it Proof}.- We will show first that any box with $\beta_{000}(P)>2$ is not broadcastable. 
Suppose by contradiction that they can be broadcast i.e. there exists a transformation $\Lambda$ such that takes a box $P_{AB}$ to  a broadcast copy $P_{ABA'B'}$.
We will use now monotonicity of anti-Robustness under linear operations that transform $LR_{ns}$ boxes into $LR_{ns}$ boxes (Observation \ref{obs:monot}).
From monotonicity and the above Lemma \ref{lem:tau-commutes} we get,
\ben 
{\bar{R}}(\tau_{rs} P_{AB}) = {\bar{R}}(P_{AB}) \le {\bar{R}}(P_{ABA'B'}) \le \label{eq:broadcasting}\\
{\bar{R}}\left [ (\tau_{rs}^{AB} \otimes \tau_{rs}^{A'B'}) (P_{ABA'B'}) \right]. 
\een
But this contradicts equation (\ref{eq:non-inc-online}). This reduction argument proves no-broadcasting of boxes  $P$ satisfying $\beta_{000}(P) > 2$.
The whole set of $2\times 2$ non-$LR_{ns}$ boxes can be written compactly as
\be
\bigcup_{r,s,t=0}^1 \{P: \beta_{rst}(P)>2\},
\ee
hence we need to have proof for 7 other values of string $rst$. We prove that if boxes with $\beta_{000}>2$ are non-broadcastable
then so are those with $\beta_{r's't'}>2$ for $r's't'\neq 000$. This is because by definition of $B_{rst}$ there is local operation which maps $B_{000}$ into $B_{r's',t'}$ and $B_{001}$ into $B_{r's'\bar{t'}}$. Hence if boxes $\alpha B_{r's't'} + (1-\alpha)B_{r's'\bar{t'}}$ were broadcastable for 
$\alpha \in [1,{3\over 4})$, then the corresponding box $\alpha B_{000} + (1-\alpha)B_{001}= B_{\alpha}$ would be broadcastable, which is disproved in section \ref{sec:lineproof}. Thus 
we have no-broadcasting on a line between $B_{r's't'}$ and $B_{r's'\bar{t'}}$ with $\beta_{r's't'}>2$. To prove this for all $\beta_{r's't'}>2$ boxes, 
we note, that reduction argument as shown above applies, with $r=r'$ $s=s'$ $t=t'$ in lemma \ref{lem:tau-commutes}. This proves the theorem. \blacksquare

\section{Conclusions}
We have shown that locality preserving operations do not broadcast $2\times 2$ non-local boxes. Moreover, this result is general since impossibility of 2-copy broadcast implies impossibility of n($>$2)-copy broadcast. Indeed, if latter were true, we could simply trace out n-2 systems and obtain 2 copies of broadcast. It is intuitive in a sense that non-locality is a resource, and it can not be brought into for free, which broadcast would do. We developed an idea of monotone in boxes paradigm, introducing anti-Robustness (or equivalently Robustness), a quantity interesting on its own. The proof uses counterintuitive property of this monotone: it does not change under irreversible operation of twirling, resembling the fact that CHSH value is preserved under twirling. In this proof we have used heavily some properties of $2\times 2$ boxes. It would be interesting to show the same for arbitrary nonlocal box, which is an open question. In fact, we consider here exact broadcasting i.e. we do not allow errors in this process. It would be interesting to prove its non-exact version as well.

\section{Appendix}

In this section we prove some results including proof of lemma \ref{lem:tau-commutes}.

\subsection{ Proof of the lemma \ref{lem:tau-commutes}}

We first prove that ${\bar{R}}(P)={\bar{R}}(\tau_{rs} P)$ with $\beta_{rst}(P)>2$. We fix values r,s,t and omit them in the following proof as 
thanks to lemma \ref{lem:CHSH} it goes the same way for all these indices.

To this end consider an arbitrary box $X\neq P$ and Y=qP+(1-q)X. Then,
\be 
\beta(Y)=q\beta(P)+(1-q)\beta(X). 
\ee
To make Y local, we need clearly $\beta(X) \leq 2$.
Let $q_0^X$ be solution of 
\be 
2=q_0^X\beta(P)+(1-q_0^X)\beta(X). 
\ee
 Let us observe that
\be
{\bar{R}}(P) = \max_X \max_{q} \{q | q P + (1-q)X \in LR\},
\ee and denote 
\be q_X(P):=\max_{q} \{q | q P + (1-q)X \in LR\},
\ee
then ${\bar{R}}(P) = \max_X q^X(P)$. Now for $\beta(P)>2, \beta(X) \le 2$, any $q > q_0^X$ there is $Y\notin LR$.
Thus $q^X(P)\le q_0^X$ for any X. 

However, we have a lemma \ref{lem:betatwo} that if $\beta(A)=2$, then $A \in LR$ (see section below). 
Hence for $q=q_0^X$, $\beta(Y)=2$, and therefore $Y\in LR$.
This implies that for any X, $q_X(P) = q_0^X$. 
Thus for $\beta(P) \geq 2$ we can equivalently write definition of anti-Robustness as
\be 
{\bar{R}}(P) = \max_X q_0^X \equiv \max_X \{q: q\beta(P)+(1-q)\beta(X)=2\}, 
\label{eq:new-def}
\ee
but we know by lemma \ref{lem:CHSH} 
that twirl of a box has same value of CHSH as that of the box for the same CHSH i.e. $\beta(P)=\beta(\tau P)$ \cite{Short}
Hence,
\be 
{\bar{R}}(P)=\max_X \{q: q\beta(\tau P)+(1-q)\beta(X)=2\}.
\ee
But according to (\ref{eq:new-def}) this is nothing but the definition of anti-Robustness of $\tau P$ i.e. ${\bar{R}}(\tau P)$. And hence ${\bar{R}}(P)={\bar{R}}(\tau P)$ for $\beta(P) > 2$. For $\beta(P)=2$ we have $\beta(\tau(P))=2$ by lemma \ref{lem:CHSH}.
Hence by lemma \ref{lem:betatwo} we have that both $P$ and $\tau(P)$ are local. It is easy to see, that for local boxes anti-Robustness is 1, hence 
the desired weak inequality.\blacksquare 

\subsection{locality of $\beta_{rst}(X) = 2$ hyperplane}
The main result of this section is the lemma below. We first show the proof of this lemma, 
and then the proof of theorem (\ref{thm:equality}) which is crucial to this proof.
{\lemma
For any $r,s,t \in \{0,1\}$ and any box X, $\beta_{rst}(X)= 2$ implies $X \in LR_{ns}$.
\label{lem:betatwo}
}

{\it Proof}
Let us fix $r,s,t$.
By theorem (\ref{thm:equality}) there is   $X \in conv\{x_0,\tilde{x}^{(rst)}_1,...,\tilde{x}^{(rst)}_n\}$ where $\tilde{x}^{(rst)}_i$ are points from the half plain defined by $\beta_{rst}(x)=2$ which belongs to ray starting at $x_0=B_{rst}$ and passing through $x_i$ which is the $i$-th of 23 (apart from $x_0$) extremal point of the set of non-signalling boxes. In other words $\tilde{x}^{(rst)}_i = p_i B_{rst} + (1-p_i) x_i$ such that $\beta_{rst}(\tilde{x}^{(rst)}_i) = 2$.
Thus $X =p_0x_0+\sum_{i=1}^np_i\tilde{x}^{(rst)}_i$. Now, since $\beta_{(rst)}(X)=2$ there is $X = \sum_i p_i \tilde{x}^{(rst)}_i$ i.e. the weight $p_0$ of $x_0$ is zero in the mixture. But it is easy to check that all $\tilde{x}^{(rst)}_i$ are local,
hence $X$ must be local itself. To see this we check that for all $r's't'$ there is 
\be
-2\leq \beta_{r's't'}(\tilde{x}^{(rst)}_i) \leq 2,
\label{eq:allbetas}
\ee i.e. 
that $\tilde{x}^{(rst)}_i$ belongs to the $LR_{ns}$ in $2\times 2$. To this end we first compute from the assumption $\beta_{rst}(\tilde{x}^{(rst)}_i)=2$ the probability $p_i$ and check for all values $r's't'\neq r,s,t$ the value of $\beta_{r's't'}$ of $\tilde{x}^{(rst)}_i$. The last check is easy if we observe that $\beta_{r's't'}(B_{rst})\in\{-4,0,4\}$ and $\beta_{r's't'}(L)\in\{-2,2\}$, where $L$ stands for any locally realistic extremal box. This holds because both nonlocal boxes $B_{rst}$ and locally realistic extremal ones $L$ can be represented (not uniquely) as vectors $v_i$ of $1$s and $-1$s (4 of them in total each corresponding to one pair of $x$ and $y$), where $1$ denotes maximal correlations of a distribution and $-1$ denotes maximal anticorrelations. Each value of $\beta_{r's't'}$ can be represented as an Euclidean scalar product of $v_i$ with again vector of 4 $1$s and $-1$s depending on sign of $\<ij\>$ in definition (\ref{eq:betarst}) where the number of $-1$ is always odd. The numbers $\{-4,-2,0,2,4\}$ follows from the fact that for each $B_{rst}$ $v_i$ has always odd number of $-1$s, and each $L$ has always even number of them.
\blacksquare

\subsubsection{geometrical theorem}

Following Bengtsson and \.Zyczkowski \cite{BengtssonZyczkowski-book}, by a cone with {\it apex} $x_0$ and some body such that $x_0\not\in $ body as a {\it base} 
we mean the set of points obtained by the following operation: taking rays (half lines) that connect $x_0$ and each point of the body. 

Thus we consider operation {\it cone} which makes cone from the body defined in the following way: $cone(x_0, conv\{x_1,...,x_n\})$ where
$x_0$ does not belong to $conv\{x_1,...,x_n\}$ and $x_1,...,x_n$ are extremal points of the body. In our case, the apex 
will be any of the maximally non-local boxes $B_{rst}$, and the body will be convex combination of other 23 extremal points of the set of non-signalling boxes.
We recall that $\beta_{rst}(B_{rst})=4$. Equality $\beta_{rst}(X)=2$ defines a hyperplane $H^{(rst)}$. By the set of $X$ satisfying $\beta_{rst}(X)\geq 2$ we mean $H_+^{(rst)}$. In what follows we fix r,s and t and omit it, as the proof goes the same way for all indices.

The main thesis of this section is the following 
{\theorem $H_+\cap B = conv (\{x_0,\tilde{x}_1,...,\tilde{x}_n\})$ where $\tilde{x}_i$ is a point from $H$ (the half plain defined by $\beta(x)=2$) which belongs to ray starting at $x_0$ and passing through $x_i$.
\label{thm:equality}
}

In what follows we use numerously the following lemma:
{\lemma If $x = a z + (1-a) x_0$ $a \in R$ and linear function $\beta(x) = \lambda \beta(z) + (1-\lambda)\beta(x_0)$, then either $\lambda = a$ or $\beta(z) = \beta(x_0)$.
\label{lem:order}
}

{\it Proof }

By linearity of $\beta$ we have  $\beta(x)= a \beta(z) + (1-a) \beta(x_0)$ but such a combination is unique in real numbers, hence either $a =\lambda$ or
$\beta(z) = \beta(x_0)$, which ends the proof.\blacksquare

In what follows, we will have $\beta(x_0)\neq \beta(z)$, but we do not state it each time.
Armed with this lemma, we can observe the following property:

{\lemma $H_+\cap B = H_+ \cap C$ where $H_+$ is a half space defined by $\beta(x)\geq 2$, $B$ is body spanned by $\{x_0,x_1,...,x_n\}$ distinct points, and $C$ is a cone obtained
by operation $cone(x_0,conv\{x_1,...,x_n\})$.
\label{lem:equality}
}

{\it Proof}

If $x \in$RHS, then $\beta(x) \geq 2$ and there exists $y \in conv\{x_1,...,x_n\}$ such that  $x = \alpha x_0 + (1-\alpha)y$ for $\alpha \in R_{\geq 0}$. By lemma \ref{lem:order}, this means that $x \in [x_0,y]$ because $\beta(x_0)=4$, $\beta(y)<2$ and $\beta(x)\geq 2$ and we have $\beta(x)\leq 4$. 
Since $\beta(x)\geq 2$, we have $x\in H_+$ which proves $x \in $LHS. Take now the converse: $x\in LHS$. This means that $x = \sum_{i=0}^n p_i x_i$, hence $x = p_0x_0 + (1-p_0) \sum_{i=1}^{n} p_i/(1-p_0) x_i$ which means $x \in cone(x_0, conv\{x_1,...,x_n\})$, and hence $x \in C$, which taking into account $x\in H_+$, gives $x\in RHS$ which proves the thesis.\blacksquare

We can now prove the following lemma, which enables us to state the main question of this section:
{\lemma H has one point of intersection with each of the segments $[x_0,x_i]$, denoted as $\tilde{x}_i$}.

{\it proof}
We have $L_i = \{x: (1-\alpha)x_0 + \alpha x_i = x\}$, $H = \{z: \beta(z)=2\}$. We want to prove that $H\cap L = \{\tilde{x}_i\}$.
To this end we observe that $x\in H\cap L$ implies $\beta(x)=2$. Taking this into account and $\beta(x_0)=4$ as well as $\beta(x_i)\leq 2$
we have by lemma \ref{lem:order} that there exists unique $\alpha \in [0,1]$ such that $x = (1-\alpha ) x_0 + \alpha x_i$, call it
$\tilde{x}_i$ i.e. $\tilde{x}_i \in [x_0,x_i]$ as we claimed.

To prove theorem \ref{thm:equality}, we first show the following inclusion:  
{\lemma $H_+\cap B \supseteq conv (\{x_0,\tilde{x}_1,...,\tilde{x}_n\})$
\label{lem:inclusion}}. 

{\it Proof}

Take $x$ from RHS. First we prove that $x \in H_+$. This is
easy since $x = \gamma_0 x_0 +\sum_{i=1}^n \gamma_i \tilde{x}_i$, by linearity of function $\beta$ we have $\beta(x)= \gamma_0\beta(x_0) + \sum_{i=1}^n\gamma_i2$ since for each $\tilde{x}_i$ we have $\beta(\tilde{x}_i)=2$. Thus, following $\beta(x_0)>2$ we have also $\beta(x) >2$ i.e. $x \in H_+$. 

We prove now that $x \in B$. To this end note that by definition $\tilde{x}_i = \alpha_i x_0 + (1-\alpha_i)x_i$, hence 
$x = (\gamma_0 + \sum_{i=1}^{n} \gamma_i\alpha_i) x_0 +\sum_{i=1}^n \gamma_i(1-\alpha_i)  x_i \in conv\{x_0,x_1,...,x_n\}$.\blacksquare

To prove the converse inclusion: $H_+\cap B \subseteq conv (\{x_0,\tilde{x}_1,...,\tilde{x}_n\})$ we need the following lemma:

{\lemma Equivalent definition of a cone C is the set of all points satisfying $x = x_0 + \sum_i \alpha_i r_i$ with $r_i = x_i - x_0$ and $\alpha_i$ are non-negative coefficients.
\label{lem:span}
}
 
{\it Proof}

We have the following chain of equivalences.  
$x = x_0 + \sum_i \alpha_i r_i$. This is if and only if  $x= x_0 +\sum_i \alpha_i (x_i-x_0)$, which is iff $\sum_i \alpha_i x_i + (1-\sum_i \alpha_i) x_0$ and this is equivalent to $\alpha \sum_i \alpha_i/\alpha x_i + (1-\alpha)x_0$ which we aimed to prove.\blacksquare  

This lemma gives the following 
{\corrolary Equivalent definition of C is the set of all points satisfying $x = x_0 + \sum_i \gamma_i \tilde{r}_i$ where $\tilde{r}_i = \tilde{x}_i - x_0$ and 
$\gamma_i$ are non-negative coefficients.
\label{col:span}
}

{\it Proof}

We know that $\tilde{x}_i = \lambda_i x_i + (1-\lambda_i)x_0$ where $\lambda_i \in R_+$ hence $\tilde{r}_i = \lambda_i(x_i - x_0) = \lambda_i r_i$, which ends
the proof, since $\tilde{r_i}$ are just scaled $r_i$ and the proof goes with similar lines to that of lemma \ref{lem:span}.\blacksquare

To complete the proof of theorem \ref{thm:equality} we now proceed with the proof of the converse inclusion: $H_+\cap B \subseteq conv (\{x_0,\tilde{x}_1,...,\tilde{x}_n\})$. Thanks to lemma \ref{lem:equality}, we may assume
that $x \in H_+\cap C$. Now, thanks to lemma \ref{lem:order}, if we take the ray with beginning $x_0$ crossing $H$ in point $\tilde{x}_0$, that passes through $x$ then if $x \in H_+$, there is $x \in [x_0,\tilde{x}_0]$. This is because $\beta(x_0)>2$ and $\beta(x_0)\geq\beta(x)>2$ while $\beta(\tilde{x}_0) = 2$. 

Hence to prove that $x \in conv(x_0,\tilde{x}_1,...,\tilde{x}_n)$, it is sufficient to show that $\tilde{x}_0$ is spanned by $\{\tilde{x}_1,...,\tilde{x}_n\}$. 
We will show it in what follows. Namely, by corollary (\ref{col:span}), there is $\tilde{x}_0= \sum_i \gamma_i \tilde{x}_i$ where $\gamma_i \in R_+$. By linearity
of $\beta$ there is $\beta(\tilde{x}_0) = \sum_i \gamma_i \beta(\tilde{x}_i)= \sum_i\gamma_i 2$. Since $\tilde{x}_0 \in H$, there is also $\beta(\tilde{x}_0) = 2$. Hence there is $\sum_i \gamma_i = 1$, which taking into account non-negativity of $\gamma_i$ shows that $\gamma_i$ forms a convex combination of $\tilde{x}_i$ which we aimed to prove. This ends the proof that $H_+\cap B = conv \{\tilde{x}_1,...,\tilde{x}_n\}$, and following lemma \ref{lem:inclusion}, ends the proof of theorem \ref{thm:equality}.

\begin{acknowledgments}
We thank T. Szarek and D. Reeb for interesting discussions. This work was supported by the Polish Ministry of Science
under Grant No. NN202231937 and later by Polish Ministry of Science and Higher Education Grant no. IdP2011 000361. It was also partially supported by EC grant QESSENCE and Foundation of Polish Science from International PhD Project: "Physics of future quantum-based information technologies".
K.H. acknowledges grant BMN nr 538-5300-0637-1.
\end{acknowledgments}
\bibliographystyle{apsrev}


\end{document}